\documentclass[preprint2]{emulateapj}

\shorttitle{AO Predictive Control}
\shortauthors{Guyon}
\usepackage{graphicx}
\usepackage[mathscr]{eucal}
\usepackage{amsfonts}
\usepackage{mathtools}
\newcommand{\matr}[1]{\mathbf{#1}}
\renewcommand{\vec}[1]{\boldsymbol{#1}}

\begin{document}

\title{Adaptive Optics Predictive Control with Empirical Orthogonal Functions (EOFs)}
\author{Olivier Guyon}

\affil{Astrobiology Center, National Institutes of Natural Sciences, 2-21-1 Osawa, Mitaka, Tokyo, JAPAN }
\affil{Steward Observatory, University of Arizona, Tucson, 933 N Cherry Ave, Tucson, AZ 85721, USA}
\affil{College of Optical Science, University of Arizona, 1630 E University Blvd, Tucson, AZ 85719, USA}
\affil{National Astronomical Observatory of Japan, Subaru Telescope, National Institutes of Natural Sciences, Hilo, HI 96720, USA}
\email{oliv.guyon@gmail.com}

\author{Jared Males}
\affil{Steward Observatory, University of Arizona, Tucson, AZ 85721, USA}

\begin{abstract}
Atmospheric wavefront prediction based on previous wavefront sensor measurements can greatly enhance the performance of adaptive optics systems. We propose an optimal linear approach based on the Empirical Orthogonal Functions (EOF) framework commonly employed for atmospheric predictions. The approach offers increased robustness and significant performance advantages over previously proposed wavefront prediction algorithms. It can be implemented as a linear pattern matching algorithm, which decomposes in real time the input (most recent wavefront sensor measurements) into a linear sum of previously encountered patterns, and uses the coefficients of this linear expansion to predict the future state. The process is robust against evolving conditions, unknown spatio-temporal correlations and non-periodic transient events, and enables multiple sensors (for example accelerometers) to contribute to the wavefront estimation. We illustrate the EOFs advantages through numerical simulations, and demonstrate filter convergence within 1 minute on a 1 kHz rate system. We show that the EOFs approach provides significant gains in high contrast imaging by simultaneously reducing residual speckle halo and producing a residual speckle halo that is spatially and temporally uncorrelated.
\end{abstract}


\keywords{Adaptive Optics --- High angular resolution --- Exoplanets}


\section{INTRODUCTION AND APPROACH}
\label{sec:intro} 

\subsection{Previous work}
Wavefront disturbances in ground-based telescope systems exhibit strong spatio-temporal correlations: Atmospheric turbulence is well-described by the frozen flow hypothesis, where discrete static turbulence screens are moving across the beam at near-constant speed and direction due to wind. On-sky wavefront sensor measurements have indeed shown that a small number of such layers accurately describe wavefront evolution \citep{Poyneer:09, 2013aoel.confE..81C}. Non-atmospheric wavefront aberrations introduced by the telescope and instrument systems are also dominated by specific patterns and modes (most frequently tip-tilt and focus modes corresponding to rigid-body misalignments) which often resonate at specific frequencies. A significant fraction of wavefront disturbances can therefore be predicted from the available recent noisy wavefront sensor measurements. Doing so can simultaneously mitigate wavefront sensor noise by averaging measurements, and eliminate the time lag due to the delay between wavefront sensor effective measurement time and wavefront correction by prediction, yielding significant improvements in adaptive optics system sensitivity and performance. 

Adaptive optics predictive controller solutions have been proposed for ground-based adpative optics \citep{doi:10.1117/12.459684, 2004SPIE.5491.1129C}. Among the solutions proposed, the Kalman filter based Linear Quadratic Gaussian (LQG) controller \citep{Paschall:93, LeRoux:04, Kulcsar:06, Looze:09} provides under conditions of turbulence statistics stationarity and unbiased gaussian measurement noise the optimal minimum residual variance solution. Recent improvements on the LQG framework provide a path to real-time implementation with currently available computing hardware. \cite{Massioni:11} proposed the distributed Kalman filter, a fast implementation relying on the frozen flow approximation. It can be deployed on large telescopes and simultaneously correct multiple atmospheric layers \citep{Gilles:13}, and can be integrated with turbulence layers identification for increased performance and robustness \citep{Massioni:15}. \cite{2014OExpr..2220894G}'s Ensemble Transform Kalman Filter (ETKF) provides a computationally efficient solution to changing atmospheric turbulence conditions. 

Feasible implementations of predictive controllers (including LQG) require the wavefront spatio-temporal evolution to be represented as a small (manageable) set of variables on which predictive filters can operate. The choice of modes (state variables in the LQG filter framework) is essential to the prediction performance, and should ideally minimize correlation between separate variables. Early predictive filter solutions \citep{Dessenne:97, Dessenne:98, Dessenne:99, 2006dies.conf..597L} operate on a single mode - usually a Zernike polynomial. Fourier modes are a natural choice under the frozen-flow hypothesis \citep{2006SPIE.6272E..2XW}. Using this representation, \cite{Poyneer:07, Poyneer:08, Poyneer:10} describe a full Fourier-based predictive control solution, which is computationally very efficient thanks to decoupling between Fourier modes and simple relationships between wind speed and Fourier mode temporal evolution. 



Predictive controllers have matured from concepts and algorithms to practical implementations and laboratory and on-sky demonstrations \citep{Petit:09} with demonstrated performance gain, and will become a core part of future AO systems. Early laboratory demonstrations and telemetry analysis show that predictive control leads to significant performance gain for vibrations control \citep{Petit:08,Poyneer:16} as well as full wavefront control \citep{2015arXiv150403686R}. \cite{2015ApOpt..54.5281C} demonstrates a $\approx$ 2 magnitude gain from predictive control using the RAVEN MOAO system telemetry on the Subaru Telescope. Implementation on AO systems is progressing from tip-tilt control \citep{2014SPIE.9148E..0OP} to more comprehensive wavefront control \citep{2014OExpr..2223565S,2016SPIE.9909E..4YS}.

\subsection{Empirical Orthogonal Functions}

In this paper, we propose to adopt the Empirical Orthogonal Functions (EOFs) as a framework for adaptive optics predictive control. EOFs are an extension of Principal Component Analysis into a multi-dimentional spatio-temporal space nominally consisting of two spatial coordinates (pupil plane coordinate) and time. The number of dimensions can be increased to also include wavelength (multi-wavelength sensing and control) and sky angle (wide field AO applications).

The EOFs approach is an extension of the matrix inversion least squares (MILS) predictors described in section 3.1 of \cite{1996ESOC...54...95L} and section 2 of \cite{2000SPIE.4007..682M}. EOFs' principal components analysis (PCA) approach to matrix inversion provides a more powerful and robust path to predictive filter derivation than the pseudo-inverse used in MILS. It also offers additional flexibility to add sensor measurements (sensor fusion) and record specific spatio-temporal patterns.

The EOFs framework originates from weather and climate forecasting. \cite{Obukhov1960} first proposed to represent the state of the atmosphere as a linear sum of orthogonal functions of spatial and temporal coordinates, empirically derived from previous measurements, to track and predict atmosphere variables. \cite{TUS:TUS1372} demonstrated that this Empirical Orthogonal Functions (EOFs) approach leads to a representation of the atmosphere with a small number of modes, from which predictions can be derived. EOFs have become a powerful technique for interpolating and forecasting meteorological data \citep{Obled1986, Aubry1991, Braud1993}. EOFs are routinely employed to predict weather patterns \citep{1989JAtS...46.3219F} as well as other complex time-variable systems, such as longer timescale atmospheric properties evolution \citep{2002GeoRL..29.1921L}, ocean temperature and salinity \citep{2003AnGeo..21..167S}, or stratospheric planetary waves \citep{2016Rolland}.

The EOFs approach explored in this paper can be described as an adaptative Karhunen-Loève Transform (KLT) in spatial and temporal coordinates: at regular time intervals, spatio-temporal KL modes are updated by decomposition of recent WFS telemetry. Prediction relies on the temporal dimension of the real-time KL modes coefficient values. A practical implementation of the EOFs approach for adaptive optics is provided in \S \ref{sec:EOFalgo}. Fundamental advantages of the approach are demonstrated by a set of example, first considering Tip-tilt prediction in \S \ref{sec:TTexamples}, and then full 2D wavefront prediction in \S \ref{sec:2DWFexamples}.

The EOFs-based approach combines several desirable features:
\begin{itemize}
\item{We show in \S \ref{sec:EOFalgo} that EOFs-based prediction converges to the {\bf global optimal linear predictive controller}. The EOFs-based approach identifies and uses all linear spatio-temporal relationships contained in the measurements. In the EOFs approach, the filter converges to the optimal linear solution with no restrictions on the input or output modes. As a result, modal or zonal representation of the input disturbances is irrelevant, as the EOFs analysis optimally identifies modes and their linear spatio-temporal relationships.}
\item{{\bf Model-free, robust}. The EOFs-based approach identifies the linear auto-regressive filter that optimally (in the least square sense) predicts future measurements with no input other than previous system telemetry and a knowledge of the loop time delay. No model and associated parameters are fed to the algorithm, as all spatio-temporal correlations that encode wavefront evolution are derived from the telemetry. Some of the previously proposed predictive control approaches use a user-specified physical model for wavefront evolution (such as the frozen flow hypothesis) and cannot recognize unexpected correlations. The EOFs-based approach does not suffer from these limitations, as no physical model of the wavefront evolution is provided, and all spatio-temporal correlations are derived from previous data. The absence of user-provided tuning parameters also makes practical operation significantly easier. The robustness of model-free approach relying purely on open-loop telemetry was previously demonstrated for tip-tilt prediction \citep{Juvenal15,2012SPIE.8447E..0ZK}; we demonstrate in this work the same benefits for larger number of modes.}
\item{{\bf Sensor Fusion}. Under the EOFs-based framework, multiple sensors can be included to participate in the wavefront estimation. Sensor providing no additional information will automatically be ignored due to lack of correlation between the input telemetry and the output solution. Sensor fusion is a core feature of EOFs-based meteorological predictive models that take multiple sensor input (temperature, wind, pressure, solar irradience, etc..) to estimate the atmosphere's future state.  We illustrate the sensor fusion capabilities of EOFs in \S \ref{ssec:sensorfusion} by combining position and acceleration inputs.}
\item{{\bf Recurring non-periodic patterns} are identified and "memorized" by the EOFs analysis. For example, an intermitent short oscillation may occur in the telescope mechanical structure, and will be automatically identified so that the predictive controller can rapidly lock on the recurring feature shortly after its onset. This capability is at the core of the EOFs-based meteorological prediction, where the last weather measurements are compared to previously encountered weather patterns. This ability, illustrated in \S \ref{ssec:transients}, is limited by Bode's integral theorem: improving the control law rejection transfer function at specific temporal frequencies decreases the rejection at other frequencies.}
\end{itemize}

We show in \S \ref{sec:HCI} that EOFs-based prediction is especially powerful for high contrast imaging.

\section{Optimal least squares solution to multivariate auto-regressive filter prediction}
\label{sec:EOFalgo}

\subsection{Notations for single variable prediction}
A wavefront sensor (WFS) is assumed here to obtain, at regular time intervals, noisy unbiased measurements of the wavefront at the entrance of an adaptive optics system. We denote a complete wavefront measurement $\tilde{\vec{w}}$ with components $\tilde{\vec{w}}_i$. We use lowercase boldface characters ($\vec{v}$) to denote vectors, with the corresponding measurement and prediction vectors denoted respectively as $\tilde{\vec{v}}$ and $\hat{\vec{v}}$. Matrices use uppercase boldface characters.

The wavefront measurements are used to perform corrections, usually performed with (a) deformable mirror(s), aimed at flattening the wavefront. Open loop control notations are adopted in this paper (the wavefront sensor does not see the corrections), noting that close loop notations can be derived from open loop notations provided that the wavefront sensor(s) and deformable mirror(s) responses are calibrated. The control loop suffers from time lag: the effect of wavefront corrections on the measured wavefront is only available after a temporal lag $\delta t$. An optimal predictive controller must simultaneously account for this time lag and average, temporally and spatially, measurements to reduce propagation of measurement noise to the residual wavefront estimation.

The $\vec{w}_i$ (coefficient $i$ of the vector $\vec{w}$ representing the full wavefront) may be the value of a wavefront mode (such at Focus) or the wavefront value over a single actuator of the deformable mirror. We note $\hat{\vec{w}}_i$ the predicted value to be computed, $m$ the number of variables in each wavefront sensor measurement $\tilde{\vec{w}}$, and $n$ the number of such measurements considered in the predictive filter (the order of the predictive filter). The $n$ previous measurements available at instant $t$ are written as a "history" vector $\vec{h}$ of $n \times m$ coefficients:
\begin{equation}
\label{equ:Ht}
\vec{h}(t) =\begin{bmatrix} \tilde{\vec{w}}_0(t) \\ \tilde{\vec{w}}_1(t) \\ \vdots \\ \tilde{\vec{w}}_{m-1}(t) \\ \tilde{\vec{w}}_0(t-dt) \\ \vdots \\ \tilde{\vec{w}}_{m-1}(t-dt) \\ \vdots \\ \tilde{\vec{w}}_{m-1}(t-(n-1)dt) \end{bmatrix}
\end{equation}

The goal of this study is to find the coefficients of the regressive filter $\matr{F}$ (written here as a 1 by $n \times m$ matrix) that provide the best estimate of a wavefront variable at time $t+\delta t$. It is assumed here without loss of generality that the variable to be predicted is one of the variables $\vec{w}_i$ measured by the sensor, and we note $\matr{F}^i$ the corresponding predictive filter:

\begin{equation}
\matr{F}^i = \begin{bmatrix} a^i_{0,0} && a^i_{1,0} && \hdots && a^i_{m-1, n-1}  \end{bmatrix}
\end{equation}
Filter coefficients are noted $a^i_{j,k}$, with indices $j$ and $k$ encoding spatial/modal and temporal dimentions respectively.

The predicted value is obtained as a linear sum of previous wavefront sensor measurements :
\begin{equation}
\label{equ:wFh}
\hat{\vec{w}}_i(t+\delta t) = \matr{F^i} \vec{h}(t) 
\end{equation}

\subsection{Training set}

The goal of this work is to find the filter $\matr{F}^i$ which minimizes the Euclidian distance between prediction and actual wavefront, temporally averaged over a sufficiently large number of measurements: 
\begin{equation}
\label{equ:min1}
min_{\matr{F}^i} < || \matr{F}^i \vec{h}(t) - \vec{w}_i(t + \delta t) ||^2 >_t
\end{equation}
Assuming that measurement errors $\tilde{\vec{w}}_i - \vec{w}_i$ are temporally uncorrelated, replacing the actual future wavefront values $\vec{w}_i$ (unknown) by the measured values yields the same optimal linear filter:
\begin{equation}
min_{\matr{F}^i} < || \matr{F}^i \vec{h}(t) -\tilde{\vec{w}}_i(t + \delta t) ||^2 >_t
\end{equation}
If the wavefront sensor temporal sampling is irregular, $\delta t$ may not be an integer multiple of $dt$, and $\tilde{\vec{w}}_i(t+\delta t)$ can be obtained by linear temporal interpolation of two nearby measurements.

The optimal filter will be derived from a training set, consisting of $l$ vectors $\vec{h}$, arranged in a $n \times m$ by $l$ data matrix 
\begin{equation}
\label{equ:dataM}
\matr{D} = \begin{bmatrix} \vec{h}(t) && \vec{h}(t-dt) && \hdots && \vec{h}(t-(l-1) dt) \end{bmatrix}
\end{equation}
and the corresponding a-posteriori measured wavefront variable values arranged in a 1 by $l$ matrix $\tilde{\matr{P}_i}$:
\begin{equation}
\tilde{\matr{P}}_i = \begin{bmatrix} \tilde{\vec{w}}_i(t+\delta t) && \hdots && \tilde{\vec{w}}_i(t + (l-1) dt +\delta t) \end{bmatrix}
\end{equation}

The algebraic representation of equation \ref{equ:min1} is
\begin{equation}
min_{\matr{F}^i}  || \matr{F}^i \matr{D} - \tilde{\matr{P}}_i ||^2
\end{equation}

By taking the transpose of the vector, the classical least-square problem is obtained:
\begin{equation}
min_{\matr{F}^i}  || \matr{D}^T {\matr{F}^i}^T - {\tilde{\matr{P}_i}}^T||^2
\end{equation}
yielding the filter solution
\begin{equation}
\label{equ:Fsol}
\matr{F}^i = \left( (\matr{D}^T)^+ {\tilde{\matr{P}}_i}^T \right)^T
\end{equation}
where $(\matr{D}^T)^+$ is the pseudo-inverse of matrix $\matr{D}^T$.

\subsection{Singular Value Decomposition of Data matrix}

The pseudo-inverse $(\matr{D}^T)^+$ in equation \ref{equ:Fsol} is computed by Singular Value Decomposition (SVD) of the data matrix, allowing for the number of singular values to be selected, and providing both numerical stability and noise filtering. Following usual notations, 
\begin{equation}
\label{equ:svd}
\matr{D}^T = \matr{U} \matr{\Sigma} \matr{V}^T
\end{equation}
and 
\begin{equation}
(\matr{D}^T)^+ = \matr{V} \matr{\Sigma}^+ \matr{U}^T.
\end{equation}

According to equation \ref{equ:Fsol}, the regressive filter $\matr{F}^i$ can then be written as 
\begin{equation}
\matr{F}^i = \tilde{\matr{P}_i} \matr{U} (\matr{\Sigma}^+)^T \matr{V}^T
\end{equation}
and the predicted value is 
\begin{equation}
\label{equ:predsol}
\hat{\vec{w}}_i(t+\delta t) = \tilde{\matr{P}_i} \matr{U} (\matr{\Sigma}^+)^T \matr{V}^T \vec{h}(t).
\end{equation}

Equation \ref{equ:predsol} provides an intuitive step-by-step interpretation of the predictive process:
\begin{itemize}
\item{$\matr{V}^T \vec{h}(t)$: The $nm$ by $nm$ matrix $\matr{V}$ contains the principal components of the data matrix $\matr{D}^T$ (see equation \ref{equ:svd}). The history vector $\vec{h}(t)$ is first mapped from the original measurement space to the data principal components space.}
\item{$\matr{\Sigma}^+ (\matr{V}^T \vec{h}(t))$: Dimensionality reduction / filtering. The dominant principal components are kept, while other components are excluded. This step selectively preserves the components of the history vector that are often encountered in the training dataset, as they hold strong predictive power. The result of this step is a $l$-length vector, refered to as the filtered history.}
\item{$\matr{U} (\matr{\Sigma}^+ \matr{V}^T \vec{h}(t))$: Projection to measurements indices. Multiplication by the $l$ by $l$ matrix $\matr{U}$ maps the filtered history to the measurements indices. The result of this mapping is a $l$-length vector, which expresses the filtered history as a linear sum of the original data matrix $\vec{h}$ vectors.}
\item{$\tilde{\matr{P}_i} (\matr{U} \matr{\Sigma}^+ \matr{V}^T \vec{h}(t))$. The filtered history, mapped onto the data matrix indices, is multiplied by the measured future values.}
\end{itemize}

\subsection{Mitigating Noise Propagation}

The least square approach may erroneously fit noise contained in the data matrix. Assuming no temporal or spatial correlations in the measurement noise, propagation of measurement noise in the filter is not a concern if the number of samples in the data matrix is significantly larger than the number of coefficients. The least-square fitting will then find the optimal coefficients that minimize noise propagation. For large size filters, it may however not be practical to ensure that the data matrix size is sufficiently large, due to computing constraints (large size SVD) or clock time to acquire the measurements. 

To solve the computation challenge and keep the size of the data matrix manageable, shorter sequences of telemetry can be processed individually, and the resulting filters $\matr{F}$ averaged. A continuous rolling average can be implemented to distribute computations in time and minimize lag between data collection and filter estimation. For example, instead of computing a filter based on a single 60 second data sequence, filters can be computed every second based on the last second of telemetry, and the last 60 such filters averaged to perform the wavefront prediction. 

A fundamental challenge to predictive control is that the statistical properties of the wavefront are continuously changing (for example, wind may be changing direction), so that the predictive filter becomes stale over time. The number of samples that should be used towards filter estimation is therefore limited, and noise propagation effects cannot be mitigated by time averaging. 
Assuming white measurement noise, the contribution of measuremement noise to the predicted wavefront state can be written from equation \ref{equ:wFh} as
\begin{equation}
{{\sigma_P}_i}^2 = \sigma^2 \sum_{m,n} \left( \matr{F}^i_{m,n}\right)^2 
\end{equation}
where $\sigma$ is the noise per measurement and $i$ is the spatial coordinate or modal index of the wavefront.
Reducing this noise term is equivalent to reducing the $L_2$ norm of the filter $\matr{F}^i$. This regularisation must be balanced against the predictive power of the filter. The Tikhonov regularization can be employed to compute a regularized filter by appending to the data matrix an identity matrix scaled by the regularization parameter $\lambda$. 
The quantity to be minimized is then
\begin{equation}
min_{\matr{F}^i}  || \matr{F^i} \matr{D} - \tilde{\matr{P}_i} ||^2 + \lambda ||\matr{F}^i||^2
\end{equation}

Equation \ref{equ:Fsol} then becomes
\begin{equation}
\label{equ:FsolR}
\matr{F}^i = \left( (\matr{D}^T_R)^+  \tilde{\matr{P}}_{i_R}^T \right)^T
\end{equation}
where $\matr{D}^T_R$ is the data matrix transpose appended with the $\lambda$-scaled identity matrix, and ${{\matr{P}_i}_R}^T$ is the a-posteriory measured vector appended with zero values.

\subsection{Extension to multivariate prediction}

As shown by equation \ref{equ:Fsol}, the computationally intensive part of the predictive filter computation, which is the SVD, is only a function of the data matrix, and is independent of the mode to be reconstructed. With little additional computation load, the predictive filter can therefore be applied to multiple wavefront variables. For example, if all wavefront variables measured by the wavefront sensor are to be predicted, the 1-by-$nm$ matrix $\matr{F^i}$ and the 1-by-$l$ marix $\tilde{\matr{P}_i}$ can be respectively replaced by the matrices:
\begin{equation}
\matr{F} = \begin{bmatrix} \matr{F}^0 \\ \vdots \\ \matr{F}^{m-1} \end{bmatrix}
\end{equation}
and
\begin{equation}
\tilde{\matr{P}} = \begin{bmatrix} \tilde{\matr{P}_0} \\ \vdots \\ \tilde{\matr{P}_{m-1}} \end{bmatrix}.
\end{equation}

The prediction equation for a $n$-order filter is then 

\begin{equation}
\label{equ:predsolg}
\hat{\vec{w}}(t+\delta t) = \tilde{\matr{P}} \matr{U} (\matr{\Sigma}^+)^T \matr{V}^T \vec{h}(t).
\end{equation}
where $\hat{\vec{w}}$ is a $m$-length vector, $\tilde{\matr{P}}$ is a $m$-by-$l$ matrix, $\matr{U}$ is a $l$-by-$l$ matrix, $\matr{\Sigma}^+$ is $l$-by-$nm$ matrix, $\matr{V}^T$ is a $nm$-by-$nm$ matrix and $\vec{h}(t)$ is a $nm$-length vector.

The prediction space (variables to be predicted) does not need to be identical to the measurement space: the process described in this section can be generalized to prediction of quantities that are not directly measured by the sensor. Conversely, this approach lends itself to sensor fusion, where the history vector $\vec{h}$ and data matrix $\matr{D}$ can include measurements from multiple sensors (for example, wavefront sensors operating at multiple wavelengths, or a combination of wavefront sensors, accelerometers and temperature sensors).

As shown in equation \ref{equ:predsolg}, the prediction is a linear function of past wavefront state estimates. In a closed loop adaptive optics control, the past wavefront state estimates are derived from both wavefront sensor measurements and deformable mirror commands. We assume in this paper that (the) wavefront sensor(s) and (the) deformable mirror(s) are linear devices, so the entire filter, from input telemetry to output predictions, is also linear. While non-linearity may affect WFS and DM, the proposed control law is linear.

\section{Example: 2D tip-tilt prediction}
\label{sec:TTexamples}

\subsection{Simulation parameters}



In this first example, a diffraction-limited near-IR tip-tilt sensor on a 8-m diameter telescope measures the position of a $m_H=9.05$ star at 1 kHz frequency in H band (1630nm central wavelength). With a 20\% system efficienty and a 0.307$\mu m$ bandpass, $N_{ph} = 6991$ photon are available per measurement. The corresponding single axis photon noise limited measurement precision $\sigma$ is for each frame:
\begin{equation}
\frac{\lambda}{\pi D \sqrt{N_{ph}}} = 0.16 mas
\end{equation}

\begin{figure*}[htb]
   \begin{center}
\includegraphics[scale=0.35]{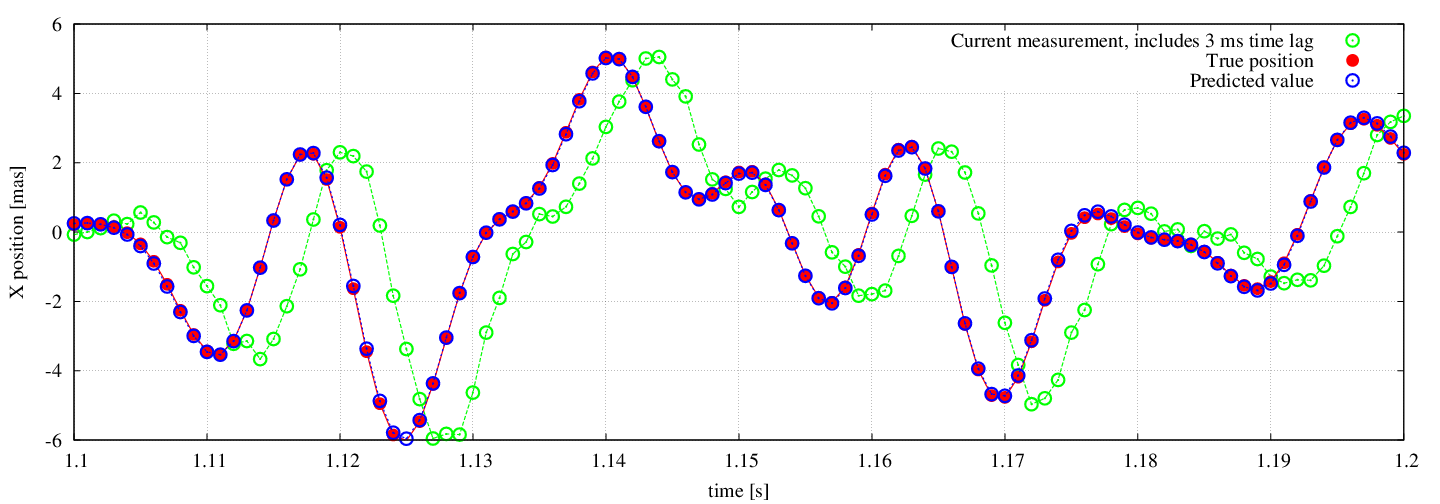} 
\begin{tabular}{cc}
{\hspace{-2cm}\includegraphics[scale=0.475]{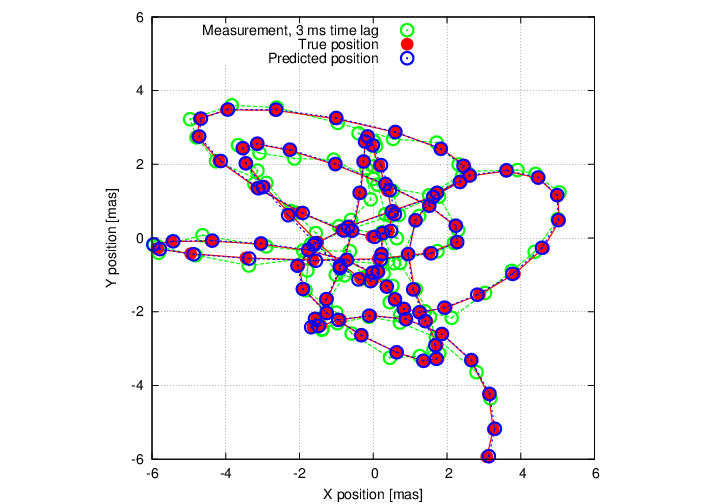}} & {\hspace{-3cm}\includegraphics[scale=0.475]{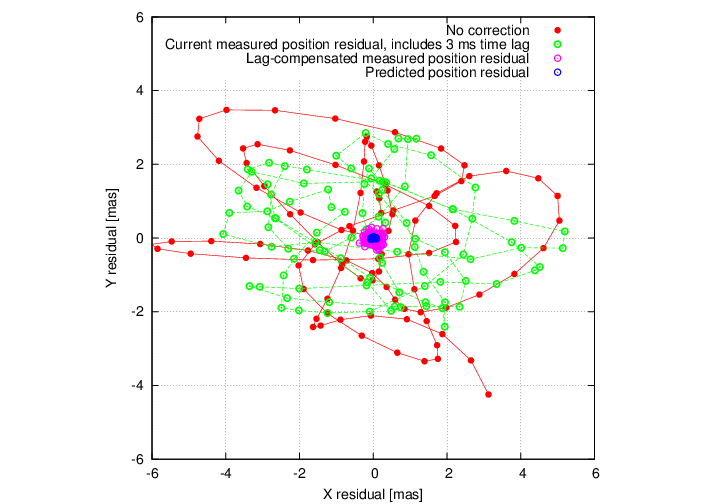}}
\end{tabular}
\end{center}
\caption[tt2D] { \label{fig:tt2D} Prediction of 2-D pointing disturbances. A 3-step lag at 1kHz sampling speed is assumed here, in the low-noise regime (measurement noise $=$ 0.16 mas). Top: Single axis (x) values, showing true pointing (red), predicted pointing (blue) and last measured position (green). Bottom left: 2-D track. Bottom right: Residual pointing error. See text for detail.}
\end{figure*} 

Measurements, true position and predicted values are shown in Figure \ref{fig:tt2D}. The input disturbance (red filled dots) is a sum of twenty individual 2D vibrations between 15 Hz and 92 Hz, each with its own amplitude, temporal phase and vibration axis (defined by a position angle). For a single vibration, the input position trajectory would be a line in the X-Y plane shown in the lower left panel of Figure \ref{fig:tt2D}. With all vibrations added, the input pointing travels along a curved line. 

\subsection{Lag compensation}

We assume a 3 $ms$ measurement lag: at any given time, the currently available measurement (green cicles in Figure \ref{fig:tt2D}) is a noisy representation of the true position 3 steps ago. For this simulation, a 60-second training set (consisting of 60,000 2D samples) was processed to produce a 800-step auto-regressive filter containing 1600 coefficients. The predictive filter is then applied on the following 60 seconds of pointing disturbances: at each time step, the last 800 measurement steps (0.8 second) are arranged as a vector which is multiplied by the 1600-by-2 element prediction matrix to yield the predicted values shown as blue circles in Figure \ref{fig:tt2D}. The top panel in Figure \ref{fig:tt2D} shows the X position for a 0.1 second section of the full 60-second sequence over which the filter was applied, and demonstrates that the predictive filter is able to closely match the input disturbance. The lower left panel shows, for the same 0.1 second section, the 2-D X/Y tracks for the input disturbance, measurement and predicted position. The close match between input disturbance and predicted values holds in the two dimensions. The filter does not treat each dimension separately, but instead identifies vibrations as two-dimensional patterns to optimally use input measurements for prediction.

\begin{figure*}[htb]
   \begin{center}
\includegraphics[scale=0.35]{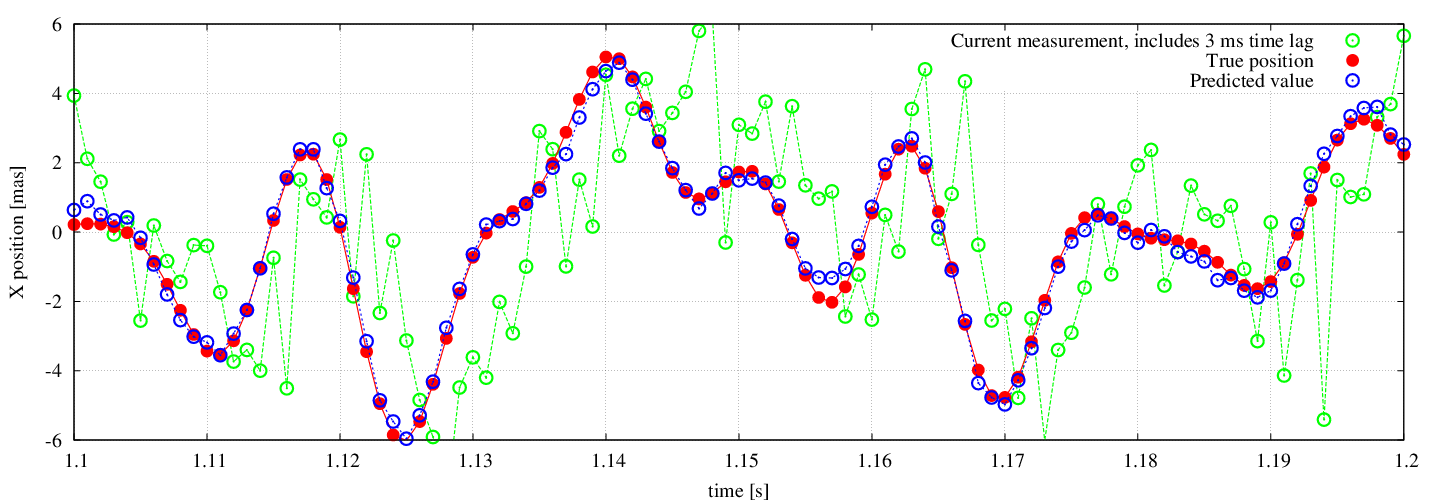} 
\begin{tabular}{cc}
{\hspace{-2cm}\includegraphics[scale=0.475]{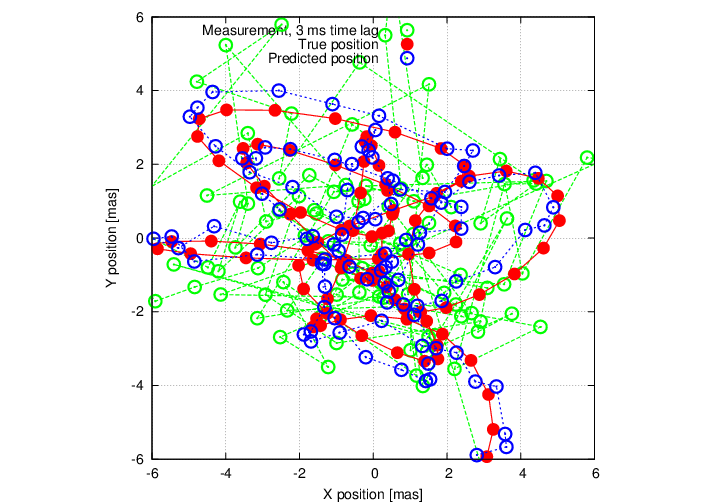}} & {\hspace{-3cm}\includegraphics[scale=0.475]{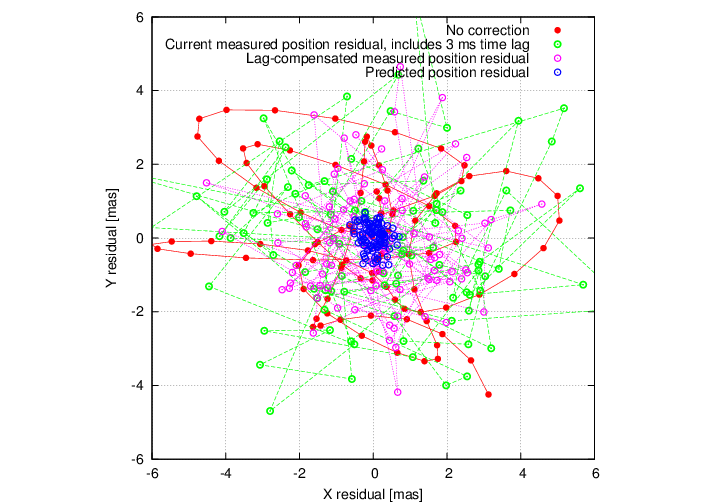}}
\end{tabular}
\end{center}
\caption[tt2D] { \label{fig:tt2Dn} Prediction of 2-D pointing disturbances in the high-noise regime. Measurement noise is 1.6 mas, 10 times larger than in the example shown in Figure \ref{fig:tt2D}. Other parameters are identical to the previsous example. Direct comparison between the two figures demonstrates the filter's ability to mitigate measurement noise while still compensating for time lag.
}
\end{figure*}

\begin{figure*}[htb]
   \begin{center}
\includegraphics[scale=0.35]{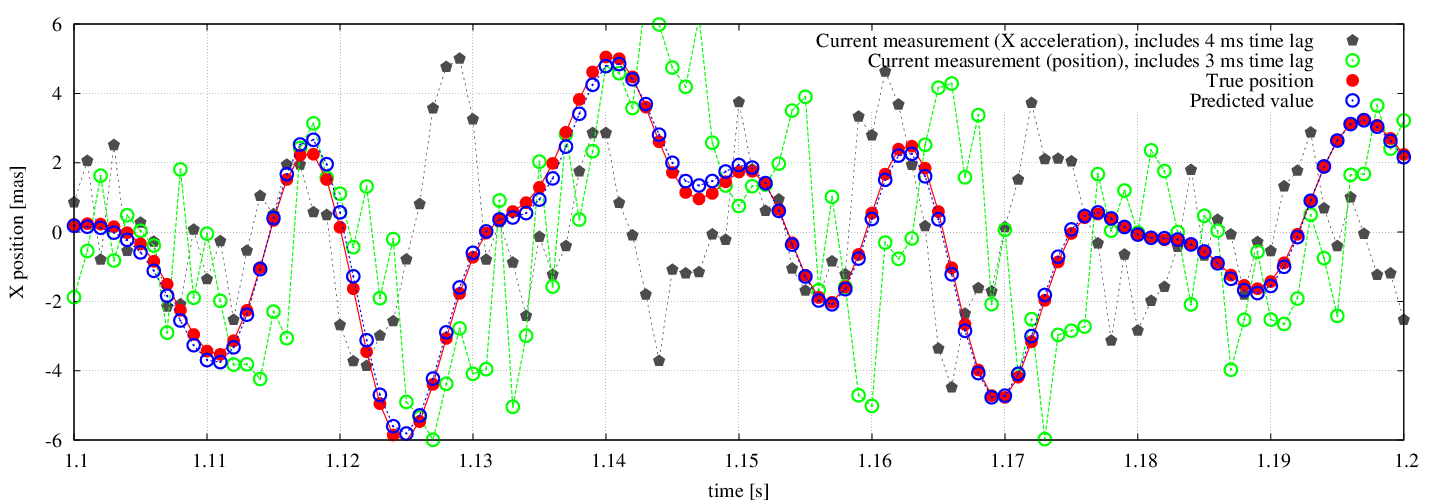} 
\begin{tabular}{cc}
{\hspace{-2cm}\includegraphics[scale=0.475]{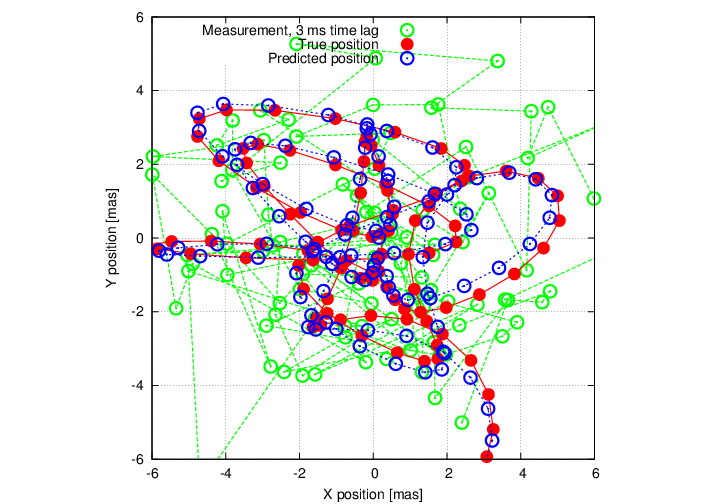}} & {\hspace{-3cm}\includegraphics[scale=0.475]{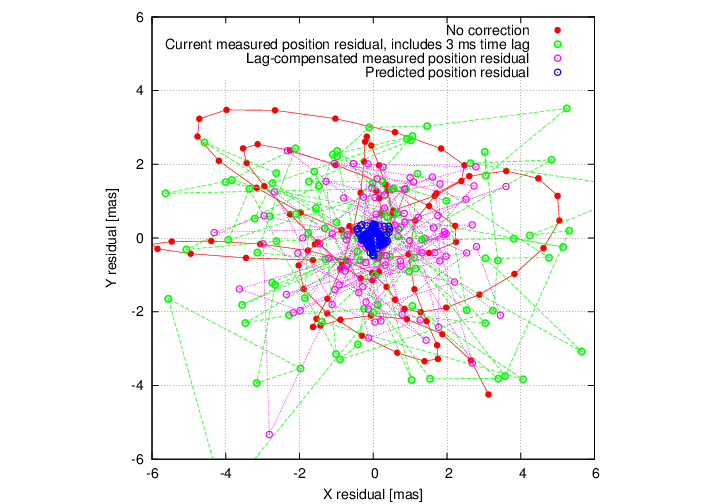}}
\end{tabular}
\end{center}
\caption[tt2D] { \label{fig:tt2Df} Prediction of 2-D pointing disturbances in the high-noise regime with accelerometer input (sensor fusion). The only difference between this example and the previous example (Figure \ref{fig:tt2Dn}) is the addition of noisy time-lagged accerometer input values in the predictive filter. Direct comparison between the two figure shows improved prediction.}
\end{figure*}

\begin{figure*}[htb]
   \begin{center}
\includegraphics[scale=0.35]{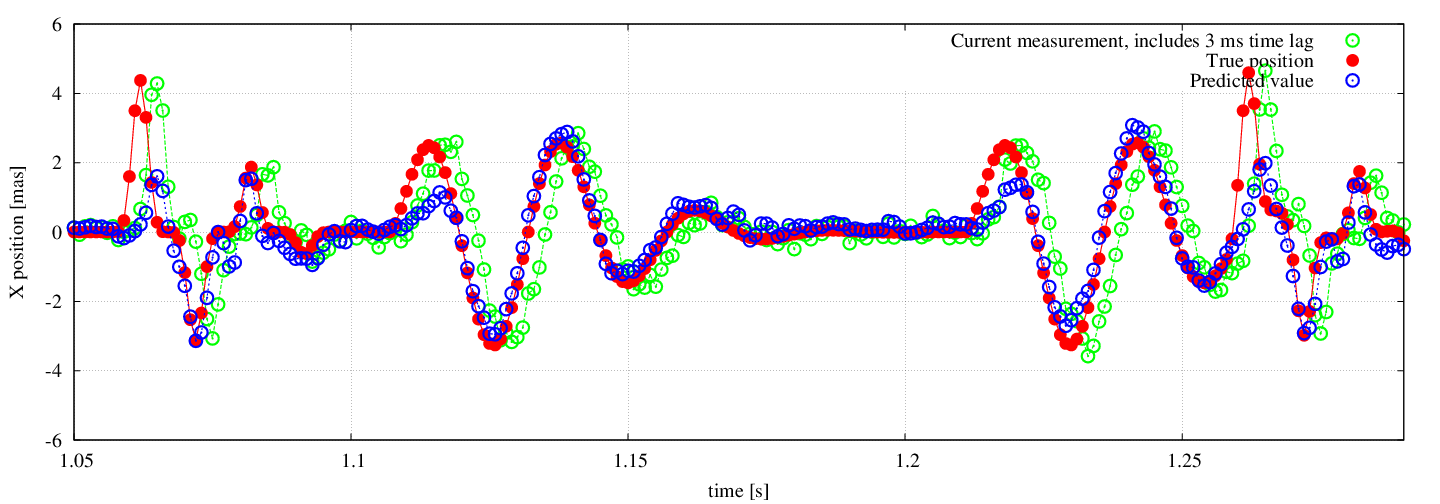} 
\includegraphics[scale=0.35]{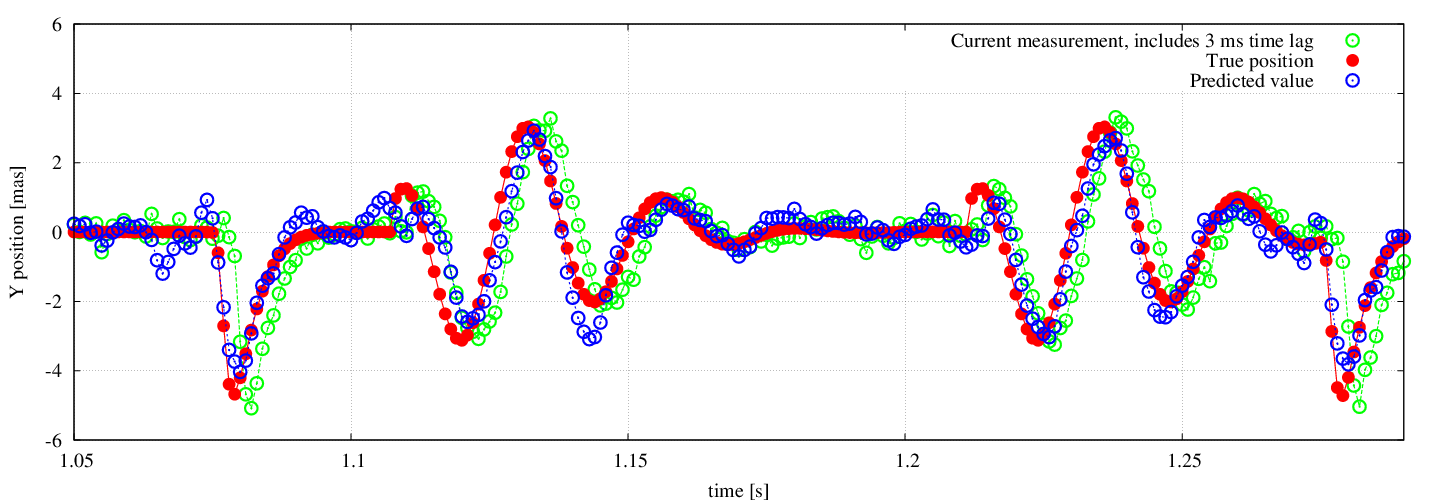} 
\includegraphics[scale=0.35]{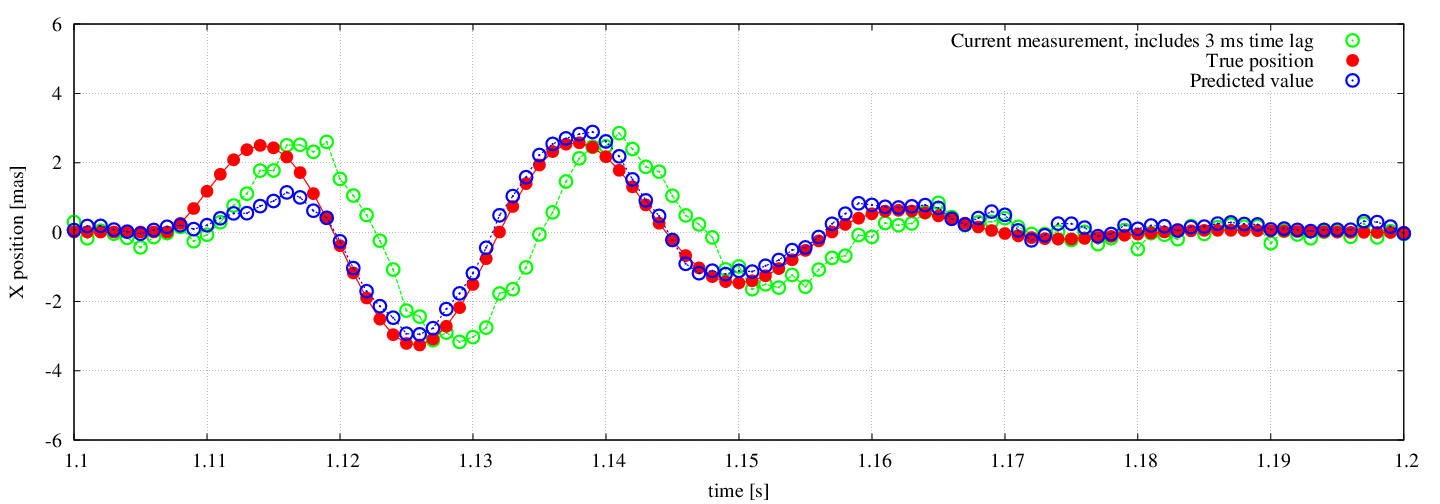} 
\end{center}
\caption[tt2Dt] { \label{fig:tt2Dt} Predictive control in the presence of non-periodic recurring patterns. Top and middle: X and Y tracks over 0.25 sec duration. Bottom: Detail, X axis, showing the response a single pattern. In this simulation, two 2D patterns occur at random times. See text for details.}
\end{figure*}

\subsection{Noise averaging}

Small differences betwen the input and measurements are visible in the lower left panel of Figure \ref{fig:tt2D}, attributable to the 0.16 $mas$ measurement noise. The predicted track matches the input significantly better than the measurements, demonstrating that in addition to compensating the measurement lag, the predictive filter is also averaging down measurement noise. The lower right panel better illustrates the combined advantages of lag compensation and noise averaging. In this 2-D representation, the residual error is shown for four scenario:
\begin{itemize}
\item{{\bf No correction (red)}. The input disturbance is left uncorrected}
\item{{\bf Correction using the last available measurement (green)}. The residual error is equal to the difference between true position and last available correction. In this low-noise example, this scenario corresponds to the best non-predictive control loop. Given the large time lag error, the residual error is only marginally better than the input disturbance.}
\item{{\bf Lag-compensated measurement residual (purple)}. If time lag were removed, and the noisy measurements were instantaneously applied as a correction, the residual error would only contain the 0.16 $mas$ measurement noise.}
\item{{\bf Predictive correction (blue)}. Correction using the predictive filter yields a 0.02 mas residual error, which is significantly better than $0.16$ mas measurement noise, demonstrating the ability to average noise.}
\end{itemize}

The predictive filter's ability to mitigate measurement noise is better demonstrated by the example shown in Figure \ref{fig:tt2Dn}, where the star brightness is $m_H=14.05$, and the corresponding measurement noise is 1.6 $mas$ per axis, 10 times larger than in the previous example. The predictive filter is able to compensate lag (top panel), and track the 2D input disturbance (bottom left). The residual error, shown as the 2D scatter of blue points in the bottom right, is 0.23 $mas$ (standard deviation measured over 60 second), or 0.16 $mas$ per axis. In the absence of input disturbances, 106 input measurements would need to be averaged to bring the noise to this level: the filter simultaneously compensates 3-step lag and averages down noise over an effective 106-step window.

This example also demonstrates the robustness and practical use of the proposed approach in the presence of measurement noise: the predictive filter is computed from a 60-second long sequence of noisy measurements, and automatically extracts patterns from noisy data. The optimal linear least-square solution adapts to noisy inputs by effectively averaging past measurement noise, without any user input. We note that in a conventional non-predictive loop, the effective averaging (encoded in the gain value for an integrator control loop) is also automatically adjusted to optimally trade lag error against noise propagation.

\begin{figure*}[htb]
   \begin{center}
\includegraphics[width=18cm]{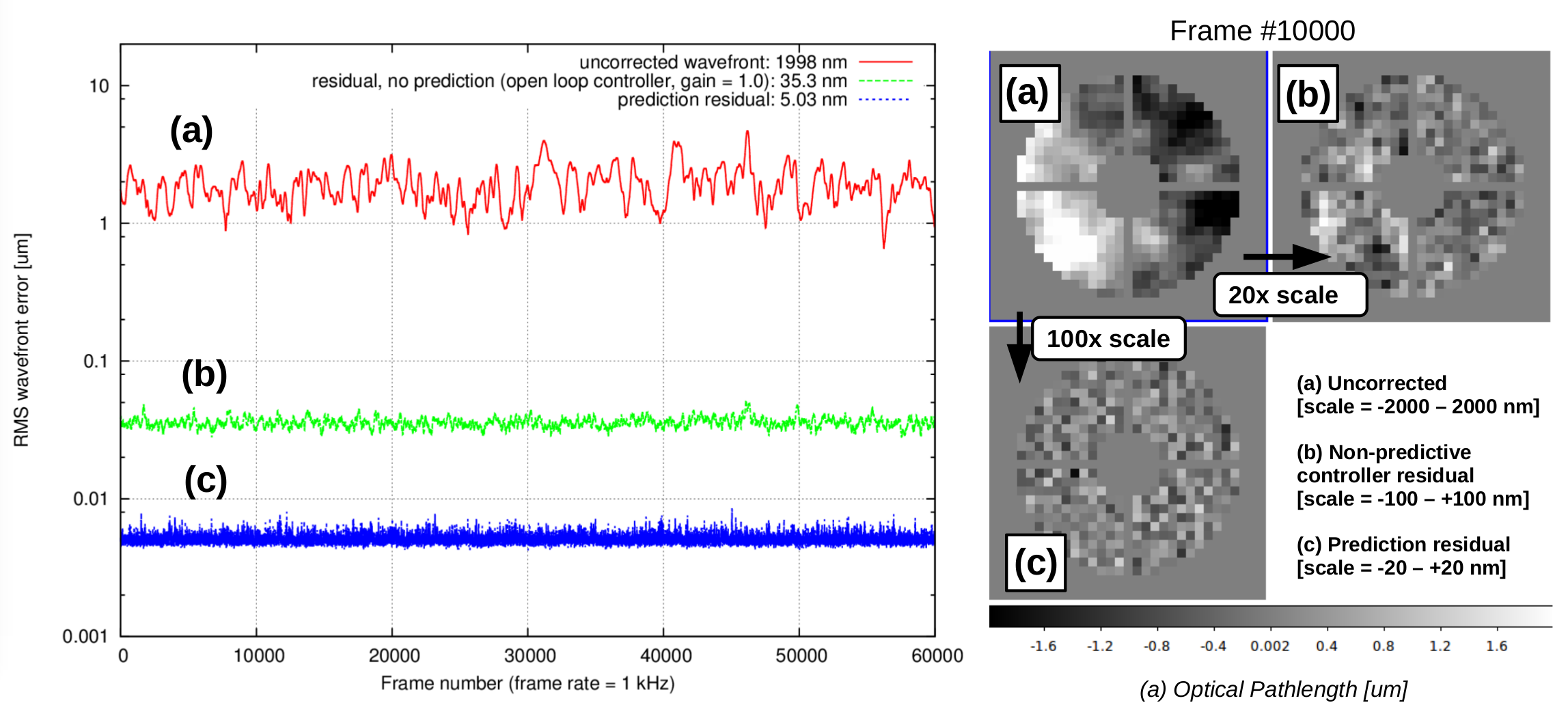} 
\end{center}
\caption{ \label{fig:atmptrace} Predictive controller performance in the bright star regime, for a 1kHz control loop with a 2-frame lag on a 8-m diameter telescope. Left: Wavefront error evolution over a 1mn time interval. Right: Sample wavefront maps.}
\end{figure*}

\subsection{Sensor Fusion}
\label{ssec:sensorfusion}

In the previous examples, the predictive filter input values were the same variables (X and Y position) as the quantities to be predicted. The EOFs framework allows for additional input variables to be included, resulting in a better prediction if the additional variables are correlated with the quantities to be predicted. Multiple sensor input can therefore be combined ({\bf sensor fusion}) to optimally exploit, for example, correlations between wavefront errors at multiple wavelengths, or between wavefronts measured on different stars in the same field. 

Figure \ref{fig:tt2Df} illustrates sensor fusion by adding accelerometer measurements to the filter input. The accelerometer provides a noisy measurement of X and Y accelerations with a 4ms delay, as shown in the top panel for the X axis. The filter is computed using a 60-second sequence of X, Y positions and accelerations, and then applied to the 4-dimension input measurements. The X and Y position measurement noise level in this example is unchanged from the previous example. Direct comparison between Figures \ref{fig:tt2Dn} and \ref{fig:tt2Df} shows that the accelerometer measurements improve the prediction from 0.23 mas residual to 0.12 mas residual. Our EOFs-based framework allows multiple sensors to be added, and does not require the sensor sampling frequencies to match: additional sensor inputs can be added in vector $\vec{h}$ (see equation \ref{equ:Ht}) without being part of the predicted wavefront variables.

The predictive filter will linearly combine noisy signals from multiple sensors to optimize, in the least square sense, the prediction. If new sensors provide redundant information, their values will be averaged with already existing measurements to reduce noise. In the EOFs-based framework discussed in this paper, the algorithm does not explicitely distiguishes between sensors: all input measurements are processed equally, and the predictive filter will choose the linear combination of input measurements that best matches the output coefficients to be predicted.

\subsection{Recurring transients patterns}
\label{ssec:transients}

A highly desirable feature of a predictive controller is to identify transient patterns occuring at random (non-periodic) times. While free atmospheric turbulence is unlikely to introduce such features, transients are often created by telescope and instrument hardware (mechanical disturbances) or interaction between wind bursts and the telescope/dome structure. Such recurring patterns are automatically identified by the EOFs approach, and become part of the predictive filter, as shown in Figure \ref{fig:tt2Dt}. In this example, the input disturbance consists of two 2D patterns occuring at random times:
\begin{itemize}
\item{Patern 1 starts with rapid positive X excursion followed by a strong negative X peak and a negative Y peak. This pattern can be seen starting at $t = 1.06$ and $t = 1.26$}
\item{Patern 2 is a 2D oscillation around a circle that expands and then contracts during a 0.1 second interval, during which approximately 2.5 revolutions are completed. This pattern can be seen starting at $t = 1.11$ and $t = 1.215$.} 
\end{itemize}

In this example, the measurement noise is set at 0.2 mas per axis, and the recurring patterns have $\approx$ 2 mas amplitude. A 200-step filter is computed using 60 second of telemetry: since the pattern durations are shorter than 0.2 second, there is no advantage extending the predictive filter size over more time steps.

The bottom panel of Figure \ref{fig:tt2Dt}, showing the $t = 1.1$ to $t = 1.2$ interval X-axis values, demonstrates the predictive filter's ability to identify and then track specific patterns. Since the patterns occur at random times, the filter is unable to anticipate its onset, and several measurements (green dots) are required before it can be identified. The pattern starts at $t = 1.108$, and rises well above the noise level by $t = 1.11$. By $t = 1.119$, pattern 2 has been identified and the prediction accurately tracks the input disturbance following the previously encountered X and Y pattern. While the bottom panel of Figure \ref{fig:tt2Dt} only illustrate the filter's pattern-locking ability for the X axis, patterns are identified and tracked in 2 dimensions.

The 2D pattern tracking behavior is more obvious on pattern 1, for which the strong rapid X excursion is left uncorrected (its onset is shorter than the time lag) but is used by the predictive controller as a signature to recognize the pattern, which is then well-tracked. 

The control law performance is still constrained by Bode's integral theorem: improvements in the rejection transfer function at specific temporal frequencies come at the expense of other frequencies. In realistic conditions where both recurring patterns and turbulence steady state are present, the filter's ability to identify and correct recurring patterns must be balanced against steady-state performance. The EOFs approach automatically finds the optimal tradeoff to minimize residual variance, provided that the training set is sufficiently long to include a representative number of recurring pattern occurrences.

The example given in this section is not representative of a realistic environment which includes both a larger number of recurring patterns and free atmosphere turbulence: in this case, the filter's ability to predict recurring patterns will be diminished. We note that some recurring patterns may slowly evolve as the telescope pointing is changing (telescope structure and components rotations), requiring frequent updates to the predictive filter (see section \S \ref{ssec:filtave}).

The EOFs-based approach's ability to identify and predict recurring patterns is especially attractive against mechanical vibrations that are near the frequency cutoff of more conventional integrator controllers. Such frequencies could be amplified by an integrator controller if they fall within the overshoot of the rejection transfer function, forcing the inegrator gain to be reduced at the expense of free atmosphere correction. With the EOFs-based approach, such vibrations can be addressed as long as their temporal frequency is within the WFS's Nyquist sampling frequency cutoff.

\section{Full wavefront estimation}
\label{sec:2DWFexamples}

\subsection{Numerical Simulation}

In this section, we apply to the EOFs-based prediction to a simulated sequence of atmospheric wavefront measurements. We consider here the bright star regime, where time lag errors dominate WFS photon noise errors, representative of an extreme adaptive optics observing a nearby star. In this regime, the optimal non-predictive controller is a open-loop integrator with unity gain: the last available WFS measurement is applied to the correcting element. In the faint star regime, a lower gain would be employed, effectively averaging multiple measurements to compute the correction.



A $m_R=4.2$ source is assumed here to be observed with a 8m diameter telescope with a 30\% linear central obstruction (45 $m^2$ collecting area) with a 0.138 $\mu m$ bandpass centered at 0.658 $\mu m$ (R-band filter). With a 20\% system efficiency, the incoming photon rate is $1.9 \times 10^9 ph.s^{-1}$. A 400-element zonal wavefront correction is assumed, with a 1 kHz wavefront sensing sampling rate, yielding 4800 photon per frame per element. The corresponding photon-noise limited measurement noise is 1.51 nm per frame, uncorrelated between frames and between sensor elements. The input wavefront error is 2 $\mu m$ RMS ($19.1 rad$ at $658 nm$), corresponding to $r_0 = 20 cm$, representative of good atmospheric conditions at major ground-based astronomical observatories.

%
%
%
%
%


Atmospheric wavefronts are simulated as a frozen flow sum of seven layers, each with a Von Karman power spectrum. The layers relative strength (relative $C_N^2$), wind speed ($m/s$), wind direction ($rad$), and outer scales (m) are listed in table \ref{tab:layers}. Spatial frequencies beyond the reach of the 400-element wavefront corrector are not included in the simulation: wavefronts are discretized into 400 phase elements corresponding to the WFS output measurements and the DM input commands. This optimistic assumption ignores how higher order aberrations contribute to Strehl ratio loss, and can corrupt the measurement of lower order aberrations through aliasing effects in the wavefront sensor.

\begin{deluxetable}{lcccc}

\tablecaption{Atmospheric turbulence layers \label{tab:layers}}
\startdata
\hline
    & Strength & Speed & Direction & Outer Scale\\
        & relative (\%) & ($m.s^{-1}$) & ($rad$) & ($m$)\\
\hline
\hline
layer 1 & 67.2 &  6.5   & 1.47  & 2\\
layer 2 & 5.1  & 6.55   & 1.57  & 20\\
layer 3 & 2.8  & 6.6    & 1.67  & 20\\
layer 4 & 10.6 & 6.7    & 1.77  & 20\\
layer 5 & 8.0  & 22     & 3.1  & 30\\
layer 6 & 5.2  & 9.5    & 3.2  & 40\\
layer 7 & 1.2  & 5.6    & 3.3  & 40
\enddata

\end{deluxetable}

\subsubsection{Non real-time computations: computing the predictive filter}
The predictive filter is computed from a 1 mn sequence ($l = $60,000 time samples) of noisy open-loop wavefront sensor telemetry, corresponding to 24,000,000 input measurements (60,000 time samples $\times$ $m = $400 degrees of freedom). A singular value decomposition of the 288,000,000-element data matrix $\matr{D}$ is then performed to compute the prediction filter. Equation \ref{equ:dataM} shows how $\matr{D}$ is assembled from the measurements: its total number of element is obtained by multiplying the number of time samples ($l = $60,000), the number of degrees of freedom ($m = $400) and the temporal order of the filter ($n = 12$ order in this example). The predictive filter computation time is dominated by the SVD computation, which requires $\approx 2 \times l \times m \times n$ floating point operations, corresponding to $\approx 3\:10^{12}$ operations: 60 GFLOPS of computing bandwidth is required to keep up with incoming telemetry. While this is well within the capacity of modern hardware (a single GPU offers $\approx$ 10 TFLOPS in single precision), predictive filter computation can be a significant challenge for higher-order systems running at multiple kHz.

\subsubsection{Real-time computations: applying the predictive filter}
The filter is then applied to the following 60 seconds, over which its performance is measured. Figure \ref{fig:atmptrace} shows, over the 1mn evaluation period, the open-loop wavefront error, and the closed loop wavefront residuals for both the integrator controller and the EOFs-based 12-step predictive controller. Each reconstruction step requires multiplication of a $400 \times 12 = 4800$ element input vector by the $4800 \times 400 = 1920000$ element prediction matrix to produce the desired 400 element solution. Assuming dense matrices, the required $\approx 4$ Gflops computing bandwidth is well within the current capabilities of conventional computers.

\subsubsection{Residual Wavefront Errors}

Figure \ref{fig:atmptrace} shows the uncorrected and corrected (with and without predictive control) RMS wavefront errors over the 60 sec evaluation period. The initial 2 $\mu$m wavefront error is reduced to 35.3 nm with an optimal non-predictive controller. The predictive controller achieves 5.03 nm thanks to time lag compensation. To achieve similar performance with a non-predictive controller, the time lag would have to be shorter than 286 $\mu$s, corresponding to a 7 kHz loop speed assuming a 2-frame delay. In the high flux regime considered here, the predictive controller relaxes the control loop speed by a factor 7. The correction is very stable in time during the 60 sec interval shown in Figure \ref{fig:atmptrace}.

Sample wavefront maps, shown on the right part of Figure \ref{fig:atmptrace}, illustrate that the predictive filter affects the spatial characteristics of the residual wavefront errors. With a non-predictive filter, the lag-dominated wavefront error exhibits low and mid-spatial frequency features which are akin to the spatial derivative the input disturbance: under the frozen flow model, spatial and temporal derivatives are interchangeable. With the predictive filter, the residual wavefront error is largely spatially uncorrelated, and is best described as "white noise" containing very little low and mid spatial frequencies. While the overall gain in wavefront error between the non-predictive and the predictive controllers is a factor 7, the performance improvement for low spatial frequencies is significantly higher.

\begin{figure}[htb]
   \begin{center}
\includegraphics[width=9cm]{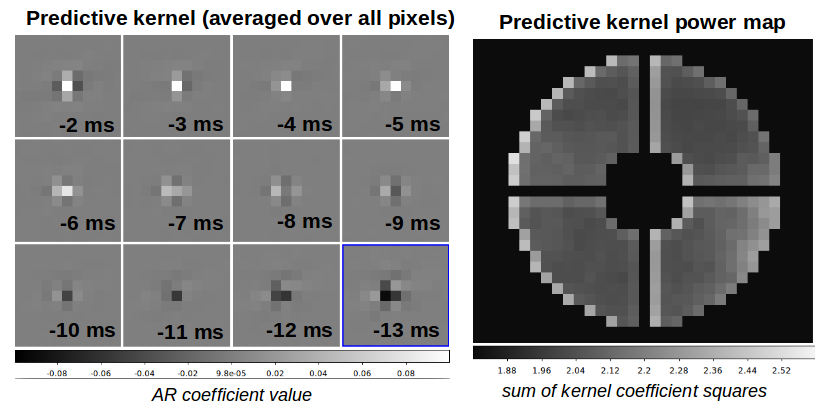} 
\end{center}
\caption{ \label{fig:kernel} Left: Zonal predictive kernel, averaged over all output actuators. Lag compensation and wind speed tracking are visible in the spatially averaged kernel, which becomes larger in size for longer time intervals. Right: Kernel power (sum of squares) as a function of actuator position on the pupil.}
\end{figure}

\begin{figure}[htb]
   \begin{center}
\includegraphics[width=9cm]{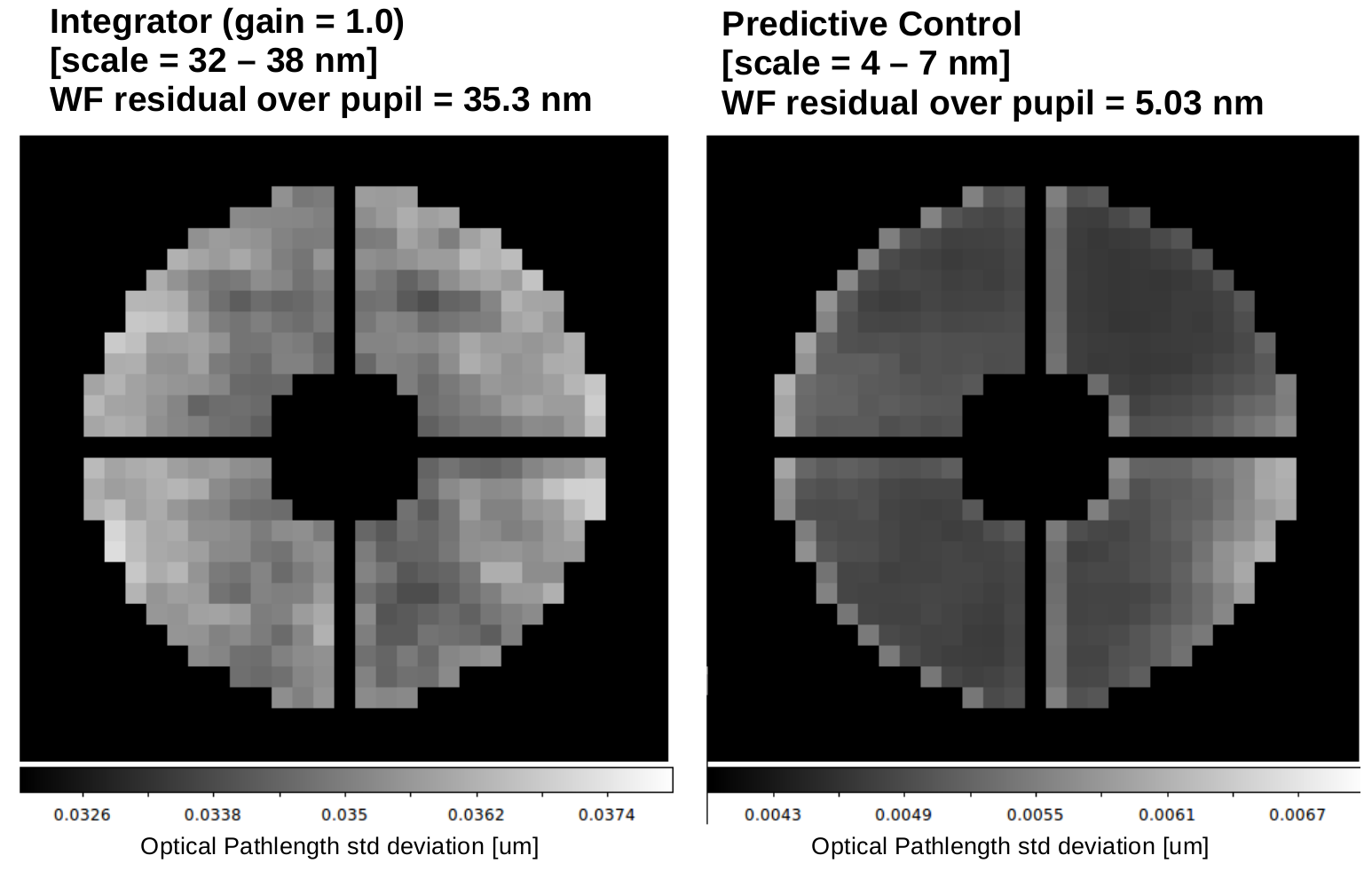} 
\end{center}
\caption{ \label{fig:stdmap} Residual wavefront error standard deviation. In the non-predictive controller (left) case, the residual error is largely independant of location on the beam. In the predictive controller (right) case, edges where dominant turbulence layers enter the beam exhibit larger residual wavefront errors. Note that brightness scales are different for both images.}
\end{figure}

\begin{figure*}[htb]
   \begin{center}
\includegraphics[width=18cm]{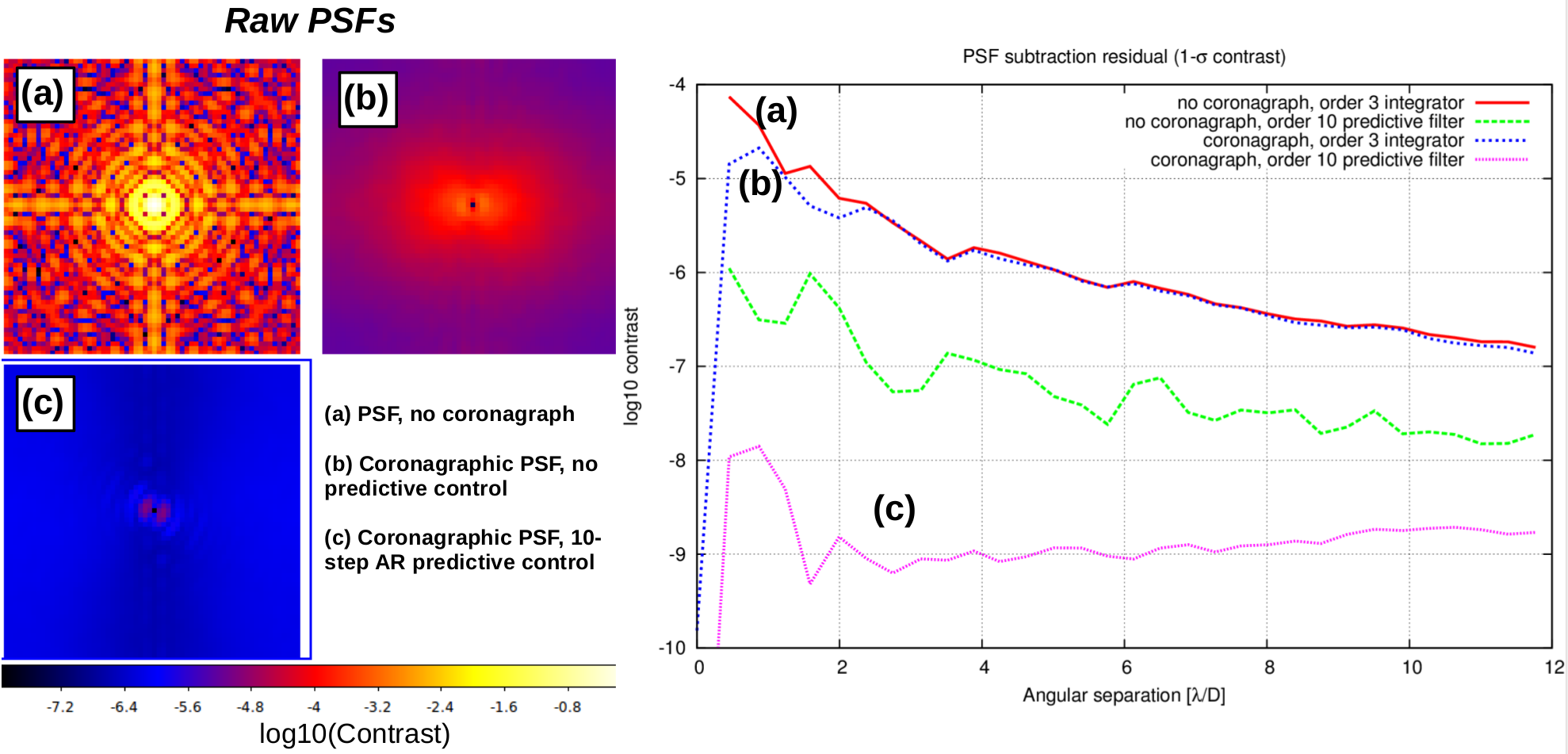} 
\end{center}
\caption{ \label{fig:contrast} High contrast imaging and predictive control. Left: Raw PSF contrast without coronagraph (a), with coronagraph + non-predictive filter (b) and with coronagraph + predictive filter (c). Right: Contrast noise (standard deviation) after PSF subtraction for a 1 hr observation sequence. Without predictive control, strong speckle noise is at the 1e-5 contrast level in the inner part of the PSF, and the coronagraph does not provide a significant advantage (small differences between curves (a) and (b)). With predictive control, speckle noise is much smaller, and the coronagraph provides a significant benefit (curve (c)) by largely eliminating the speckle-pinning effect (amplification of speckle noise by PSF diffraction features). All images and contrast values are computed at $\lambda =$ 2.11 $\mu m$ without photon noise (only speckle noise is included). The R-band wavefront sensor measurements include photon noise. See text for details.}
\end{figure*}

\subsection{Predictive Kernel}

There are 5 dimensions to the predictive matrix: input measurement position on the pupil (2D), output estimation position on the pupil (2D), and lookback time. In Figure \ref{fig:kernel}, we inspect the predictive matrix by projection and averaging on a subset of dimensions. For each output wavefront pixel, we can write the predicted value as the product between a 3-D predictive kernel and 3-D image cube containing previously measured wavefront maps. The 3-D kernel then shows how each previous 2-D wavefront measurement contributes to the single output pixel value. In the left part of Figure \ref{fig:kernel}, we have averaged all 3-D kernel after re-centering the input wavefront pixel coordinate on the output pixel coordinate. Each of the 12 time slice of the average kernel is shown, ranging from a 2-frame delay (the most recent measurement) to a 13-frame delay. As expected, the strongest coefficient is the central pixel of the first slice: the most recent WFS measurement of the output pixel provides a good estimate. 

\subsubsection{Time lag compensation}
The central pixel value becomes smaller, and ultimately negative, as lookback time increases. Considering temporal evolution only, and assuming that over short timescales the wavefront pixel value is a linear function of time, this positive-negative feature compensates for time lag: the temporal derivative (oldest values subtracted from recent values) is added to the recent values to extrapolate the measurements to a future time.

\subsubsection{Wind tracking}
Over the 13 ms time delay between the kernel's last slice and the prediction time, the fastest layer's 22 $m.s^{-1}$ wind speed corresponds to a 0.9 WFS element translation. The kernel therefore expands by approximately 1 pixel around the origin as lookback time increases. In the fastest layer (angle = 3.1 rad), wind pushes turbulence from left to right in the figure's coordinates: this corresponds to the positive (white) values on the left of the central pixel. Other (slower) layers contribute to the complex kernel shape, along with spatial correlations in the Kolmogorov turbulence spatial spectrum. 

\subsubsection{Kernel spatial variations}
The right part of Figure \ref{fig:kernel} shows the kernel power (sum of squared kernel coefficients) as a function of spatial location on the pupil. We note that the sum of all kernel coefficients is unity (the prediction must be correct for a static wavefront), so the kernel power tends to increase when a small number of measurements contribute to the prediction. The top and left edges of the pupil corresponds to incoming wind directions, and clearly show that for these pixels, there is a lack of previous measurements, so the optimal estimate is derived from a smaller number of more recent measurements. The effective averaging timescale is optimally adapted to location and wind speeds. Correspondingly, residual wavefront errors are larger at the edges corresponding to incoming turbulence (see Figure \ref{fig:stdmap}). An undersized pupil mask may be introduced to yield a slight wavefront quality improvement if the corresponding loss in flux and angular resolution is acceptable.

\section{High contrast imaging}
\label{sec:HCI}

To enable high contrast imaging, the AO system must minimize starlight intensity at small angular separation from the central star, and the residual speckle halo must be as smooth as possible for PSF substraction. 

Predictive control has long been recognized as essential for high contrast imaging \citep{1995ApJ...454L.153S}. With a non-predictive controller, time lag creates bright long-lived speckles that do not average quickly in long exposures. This "speckle noise" is the main limit to high contrast imaging performance. Predictive control is key to push raw contrast beyond the limit imposed by incoming photon rate \citep{2005ApJ...629..592G}, and to enhance PSF calibration performance by eliminating the slow speckles.

\begin{figure}[htb]
   \begin{center}
\includegraphics[width=9cm]{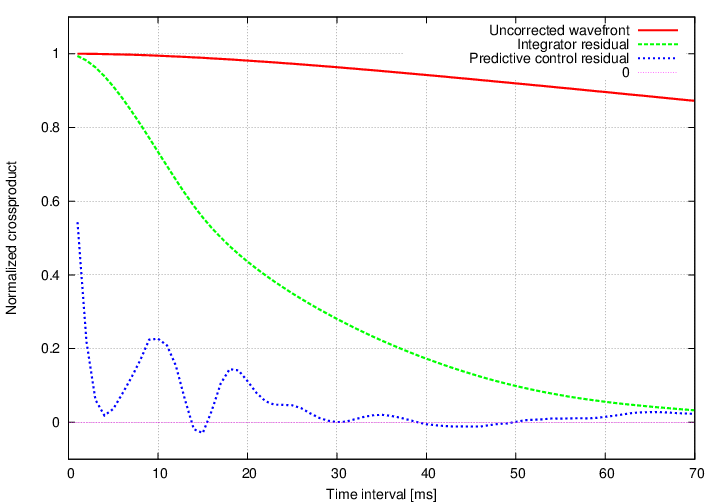} 
\end{center}
\caption{ \label{fig:WFcorr} Correlation (normalized cross-product) between wavefronts as a function of time interval.}
\end{figure}

The 1mn-long wavefront sequence computed with and without the predictive filter is fed to an idealized coronagraph that perfectly removes the on-axis unperturbed point source component of the complex amplitude field. Raw contrast images, computed at $\lambda =$ 2.11 $\mu m$, are shown in Figure \ref{fig:contrast} (left): the predictive filter reduces speckle halo intensity by $\approx$ 100$\times$ at small angular separations. The residual speckle halo is both fainter and smoother with predictive control. The nearly horizontal direction of the fastest turbulent layer yields an asymetric speckle halo which is brighter along the horizontal direction. The predictive controller gain is largest at small angular separations: while the speckle halo intensity is larger near the center in the non-predictive controller case, it becomes nearly independent of angular separation with predictive control. 

Residual noise after PSF calibration is derived by differencing two long exposure PSFs, corresponding to the first and second halves of the 60 second wavefront sequence. Assuming no wavefront error correlation beyond 60 second, the residual speckle noise is scaled by $\sqrt{60/3600} = 0.13$ for a 1 hr observation time. The $1 \sigma$ speckle noise level for a 1hr observation is shown in Figure \ref{fig:contrast} (right). 

\subsubsection{Non-predictive control}
The non-predictive control cases shows a $10^{-5}$ contrast speckle noise at 2 $\lambda/D$, which corresponds to a $10^4$ detection contrast assuming a 10-$\sigma$ threshold. The detection limit is 10 times better at 6 $\lambda/D$. The coronagraph does not improve detection limits in this atmospheric speckles dominated regime. While our assumptions are highly optimistic (bias-free photon-noise limited wavefront sensing, lack of non-common path errors, wavefront achromaticity), we note that PSF subtraction by angular differential imaging on current extreme-AO systems is already approaching this limit. For example \cite{2016A&A...587A..57Z} report a $4 \times 10^{-6}$ contrast K1-band photometric error of the HR8799e planet at 7.5 $\lambda/D$ separation, which is approximately 10 times above the curves on Figure \ref{fig:contrast}. 

\subsubsection{Predictive control}
With predictive control, residual speckles are fainter than the static non-coronagraphic PSF, and are therefore amplified without a coronagraph \citep{2004ApJ...612L..85A, 2007ApJ...669..642S}: Figure \ref{fig:contrast} does indeed show a significant contrast improvement from the coronagraph. The with-coronagraph contrast gain between non-predictive and predictive control is approaching 4 orders of magnitude at 2 $\lambda/D$, which is only partially explained by raw contrast ($\approx$ 100 - 1000 $\times$ gain). Better averaging of speckles into a smooth halo brings another $> 10\times$ contrast detection gain. As shown in Figure \ref{fig:WFcorr}, the residual wavefront decorrelation time is $\approx$ 10$\times$ shorter with a predictive controller than it is with an integrator.

\section{Conclusions}

Empirical Orthogonal Functions (EOFs) provide a powerful yet flexible implementation of linear optimal predictive control. EOF prediction combines several highly desirable features: sensor fusion, pattern matching for transient events, fully automatic model-independent tuning, ability to identify and use any linear spatio-temporal relationship and robustness against sensor noise. The EOF implementation described in this paper converges to the optimal linear auto-regressive filter without requiring a physical model of the wavefront temporal evolution, and does not require user input/tuning. 

The examples provided in this paper demonstrate the EOF approach high performance and fast convergence. A 7 $\times$ improvement in residual wavefront error is demonstrated in \S  \ref{sec:2DWFexamples}, and several order of magnitudes gain in detection contrast are obtained in \S \ref{sec:HCI}. We will address the use of optimal linear prediction in closed-loop modal control for high contrast imaging in our forthcoming paper: Males and Guyon (2017, in prep.).

\subsection{Hardware considerations}

The EOF-based approach presented in this paper predicts future sensor measurements in an open loop configuration, so it can be integrated in a close loop control architecture provided that pseudo-open loop telemetry can be reconstructed as in the optimal modal control of \cite{1994A&A...291..337G,1995A&AS..111..153G}. 

The technique therefore requires reliable open loop wavefront estimates to be constructed from real-time wavefront sensor(s) and deformable mirror(s) telemetry. Knowledge of hardware/software time lags, deformable mirror and wavefront sensor responses are therefore required for successful implementation, and calibration errors will lead to poor performance and possible instabilities. While uncorrelated WFS measurement noise is properly averaged by the filter, non-linear responses of the WFS and DM have not been addressed in this paper, and can limit performance. These calibrations errors are fundamental limits to predictive control techniques, and not unique to the EOF-based approach.

Unlike predictive approaches relying on turbulence models, the technique does not explicitely predict wavefronts, but only predicts future wavefront measurements, so close loop control will only address the sensor's measurement space. This can be a limitation in wide-field AO architecture where the wavefront should be corrected along a sky direction for which there is no WFS: in this case, a sensor should be deployed in the desired sky location to train the predictive filter control, or EOFs must be combined with a turbulence model.

More importantly, because the predictive controller is optimized to reproduce future WFS measurements by linear combination of past WFS measurements, correlated noise and calibration errors in the WFS will drive the reconstruction away from the optimal solution. For example, a periodic additive noise in the WFS will be misinterpreted as a linear relationship between past and future wavefront states, adding errors in the predicted wavefronts. In addition to WFS correlated noises, the reconstructed pseudo open loop telemetry that serves as the input to the EOFs-based algorithm may contain correlated noises due to timing errors, WFS and DM non-linearities. While exploring the full range of such behaviors and their effect on performance are beyond the scope of this study, they should be taken into consideration when implementing and deploying the technique on real hardware.

\subsection{Wavefront statistics and Filter averaging}
\label{ssec:filtave}

The EOF-based approach relies on wavefront statistics stationarity for optimal performance. Deviations from stationarity will result in poor performance and possible unstable behavior: changing conditions (variable atmospheric seeing, wind speed and direction) will render the predictive filter stale, and potentially counter-productive. The predictive filter should therefore be updated sufficiently frequently to adapt to continuously changing conditions.

However, because the EOF implementation presented in this paper is not constrained by physical models, the predictive filter contains a large number of degrees of freedom, and can only be reliably derived with a sufficiently large number of measurements. In a regime where the number of measurements is small (or very noisy), the EOF process may artifically fit measurement noise in the training set. This can be mitigated with hybrid approaches, combining EOF with a physical model of wavefront evolution to constrain the predictive filter.

Additionally, the training set used to derive a filter should be sufficiently long to experience the spatial and temporal patterns that are likely to occur during predictive control. For example, in the presence of recurring disturbance patterns, the effective duration of the training set should be chosen to include multiple instances of the patterns.

Understanding the tradeoff between filter noise averaging and filter agility to changing conditions is essential to achieving high performance. The numerical simulations presented in this paper show that a filter computed from 1 mn training set is sufficient to yield significant gains in the example considered, so the approach should be robust to slowly changing atmospheric conditions. For fainter stars (higher WFS noise), the filter averaging time could become sufficiently long for the wavefront statistics non-stationarity to significant affect performance.

\acknowledgments    

The authors thank Michael Fitzgerald, Caroline Kulc{\'s}ar and Henri-Fran\c{c}ois Raynaud for their valuable suggestions to improve the manuscript. Linear algebra computations for this work were performed with the MAGMA library \citep{tdb10,tnld10,dghklty14,ntd10,ntd10_vecpar}.


\bibliography{ms}

\end{document}